# Temperature and number evolution of cold cesium atoms inside a wall-coated glass cell[*]

Huang Jia-Qiang[a)c)], Zhang Jian-Wei[b)c)†], Wang Shi-Guang[b)c)],

Wang Zheng-Bo[b)c)], and Wang Li-Jun[a)b)c) ††]

[a)]Department of Physics, Tsinghua University, Beijing 100084, P. R. China
[b)]Department of Precision Instruments, Tsinghua University, Beijing 100084, P. R. China
[c)]Joint Institute for Measurement Science (JMI), Tsinghua University, Beijing 100084, P. R. China

We report an experimental study on the temperature and number evolution of cold cesium atoms diffusively cooled inside a wall-coated glass cell by measuring the absorption profile of the $6^2S_{1/2}$ (F=4)→$6^2P_{3/2}$ (F'=5) transition line with a weak probe laser in the evolution process. We found that the temperature of the cold atoms first gradually decreases from 16 mK to 9 mK, and then rapidly increases. The number of cold atoms first declines slowly from $2.1 \times 10^9$ to $3.7 \times 10^8$ and then falls drastically. A theoretical model for the number evolution is built and includes the instantaneous temperature of the cold atoms and a fraction p, which represents the part of cold cesium atoms elastically reflected by the coated cell wall. The theory is overall in good agreement with the experimental result and a nonzero value is obtained for the fraction p, which indicates that the cold cesium atoms are not all heated to the ambient temperature by a single collision with the coated cell wall. These results can provide helpful insight for precision measurements based on diffuse laser cooling.

## 1. Introduction

Diffuse laser cooling is a convenient technique widely applied in the field of cold atomic physics. It was first proposed by Y. Z. Wang in 1979[1], and subsequently demonstrated in atomic beam slowing experiments by several groups[2, 3]. In this method, a diffuse light field is implemented to automatically compensate the varying Doppler shift of moving atoms by absorbing photons from the proper incidence angle that satisfies the resonant condition. In practice, an integrating sphere is commonly used to generate the diffuse light field. With its robust cooling efficiency and simple setup, diffuse laser cooling has been a favorable scheme for building cold atomic clocks. The first attempt in applying diffuse laser cooling in cold atomic clocks was carried out by Guillemot et al. in 1997[4]. They successfully realized cold atoms in a high-vacuum cesium vapor cell surrounded by a spectralon integrating sphere. Later, the spectralon integrating sphere was replaced with a reflecting copper cylinder[5]. Based on this setup, a compact cold cesium atomic clock, called HORACE, was developed[6]. Meanwhile, a cold rubidium atomic clock using diffuse laser cooling was demonstrated by Cheng et al. in 2009[7,8].

With a narrow velocity distribution, the cold atoms inside an integrating sphere form a desirable sample for atomic clocks or for other spectroscopy experiments. Nevertheless, because of heating during collisions with the glass cell wall, the free cold atoms can only exist for a finite lifetime after the cooling laser light is blocked. This finite lifetime determines the Ramsey interval, as well as the frequency stability limitation of such atomic clocks. For both cesium and rubidium, the lifetime of cold atoms were reported to be approximately 80 ms[9, 10]. The number evolution of the cold rubidium atom was measured previously [10]. However, the theoretical analysis did not fully explain the experimental result because the variation of the atomic temperature over time was not included in that model. Moreover, the temperature evolution was, to our knowledge, never reported.

A further study of the temperature and number evolution of the cold atoms in a cell is indispensable, not only to understand the fundamental physics of the behaviors of cold atoms in the cell, but also to provide helpful guidance in the special applications based on the cold atoms in the cell, for instance, to optimize the Ramsey pulse sequence for the atomic clock. In this paper, we report a detailed study of the temperature and number evolutions of the cold cesium atoms loaded by diffusive laser-cooling inside an integrating sphere, and provide a theoretical model to explain the two evolutions. The theoretical model agrees very well with the experimental data.

## 2. Experimental setup

The experimental setup is shown in Fig. 1. The cooling laser and the repumping laser are both diode lasers, and locked using the standard saturation absorption (SA) techniques. The cooling laser is locked to the cycling transition $6^2S_{1/2}$ (F=4)→$6^2P_{3/2}$ (F'=5) with a red detuning of 10 MHz. The repumping laser is locked to the transition $6^2S_{1/2}$ (F=3)→$6^2P_{3/2}$ (F'=4), to prevent cesium atoms from accumulating in the $6^2S_{1/2}$(F=3) level. The two lasers are combined into one beam by a polarization beam splitter (PBS), and coupled into a multimode optical fiber. The fiber is then divided into four output fibers. Fixed on the horizontal median plane of the integrating sphere, the four output fibers inject the combined laser light along two perpendicular axes. The internal diameter of the integrating sphere is 50 mm. The inner surface of the sphere was sprayed with a white reflective coating of 98% reflectivity at a 852-nm wavelength. A 40-mm diameter Pyrex vapor cell containing cesium is placed symmetrically inside the sphere. The inside wall of the glass cell is coated with the alkene material, Alpha Olefin Fraction C20-24, using the procedure described in Ref. [11-12]. The key properties of an antirelaxation coating are the absence of free electron spins, which could cause alkali-metal spin destruction, and a low dielectric constant, which could result in a low adsorption energy. These properties give alkali-metal atoms not only a long relaxation time[11-12], but also an enhanced collection efficiency in laser cooling[13-14]. A glass tube connected to the bottom of the cell links the bulb to a cesium reservoir and an ion pump. The temperature of the cesium reservoir is kept at -5 ºC and the vacuum is maintained at 2.3×10$^{-9}$ mbar by the ion pump.

---
[*]Project supported by the National Natural Science Foundation of China (11304177).
[†]Corresponding author. E-mail: Zhangjw@tsinghua.edu.cn
[††]Corresponding author. E-mail: lwan@tsinghua.edu.cn

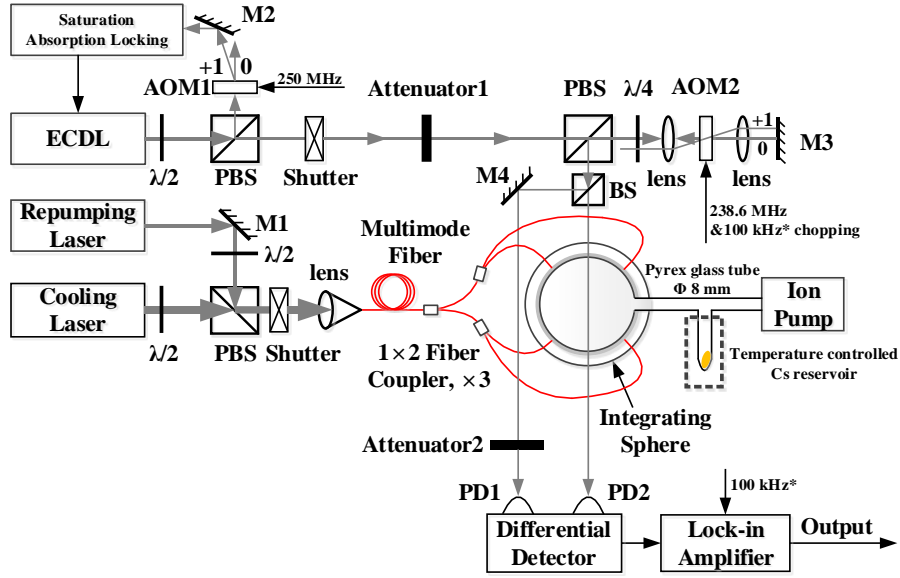

**Fig. 1.** Schematic of the experimental setup. ECDL: extended-cavity laser diode; AOM1, AOM2: acoustic-optical modulator; PBS: polarization beam splitter; BS: beam splitter; λ/2, λ/4: wave plate; M1, M2, M3, M4: mirror; PD1, PD2: photon detector.

An extended-cavity diode laser (ECDL), with a linewidth of 600 kHz, is used to probe the cold cesium atoms. A small part of the ECDL light is blue detuned 250 MHz by an acoustic optical modulator (AOM1), and the frequency is locked to the crossover transition of $6^2S_{1/2}$ (F=4)→$6^2P_{3/2}$ (F'=3) and $6^2S_{1/2}$ (F=4)→$6^2P_{3/2}$ (F'=5). The other part of the ECDL is blue detuned 477.2 MHz by an AOM2, and is resonant with the transition of $6^2S_{1/2}$ (F=4) →$6^2P_{3/2}$ (F'=5). The power of the probe laser is 0.3 μW, and the beam section is about 0.03 cm$^2$. The probe laser frequency is scanned by changing the driving frequency of the AOM2 during measurements. This laser beam is split into two beams of which one beam passes through the integrating sphere along the vertical direction via two 2 mm holes, and the other travels outside. The absorption signal is measured by a differential detector to suppress the common-mode amplitude noise of the probe laser. In order to further improve the signal-to-noise-ratio (SNR), the probe laser is chopped cut off at 100 kHz by switching the RF signal on AOM2, and the absorption signal is processed by a phase-lock loop.

### 3. Results and discussion

The three hyperfine spectral lines from the $6^2S_{1/2}$ (F=4) to $6^2P_{3/2}$ (F'=3, 4, 5) level transition of the cold cesium atoms inside the glass cell can be clearly detected, as shown in Fig. 2. The hyperfine spectral line shape P(ω) can be written as

$$P(\omega-\omega_a) = A \cdot V(\omega-\omega_a), \qquad (1)$$

where A is an area factor, $\omega_a$ is the resonance frequency and $V(\omega-\omega_a)$ is the normalized Voigt profile. In the normalized Voigt profile expression, the Gaussian linewidth, $\Delta\omega_G$, can be expressed as

$$\Delta\omega_G = 7.163\times 10^{-7} \omega_a \sqrt{T/M}, \qquad (2)$$

where T is the temperature of the cold atoms and M is the cesium atomic mass. Thus, the temperature T can be obtained if $\Delta\omega_G$ is known. The number of cold atoms can be deduced from the area factor[15-16] A, which can be written as

$$A = \pi r_e cf \cdot \Delta N_L \cdot I(\omega), \qquad (3)$$

where $r_e$=2.82×10$^{-15}$ m is the classical electron radius, c the speed of light, $f_e$=0.437 the oscillator strength, $\Delta N_L$ the population difference between level $6^2S_{1/2}$ (F=4) and level $6^2P_{3/2}$ (F'=5) within the probe laser beam and $I(\omega)$ the laser spectral intensity. By fitting a Voigt curve to the 4-5' hyperfine spectral line in Fig. 2, the Gaussian linewidth is obtained as (2.71±0.11) MHz, and the population difference within the probe laser beam is $\Delta N_L$=(8.01±0.06)×$10^6$. The calculated temperature of the cold cesium atoms is (16.0±1.3) mK.

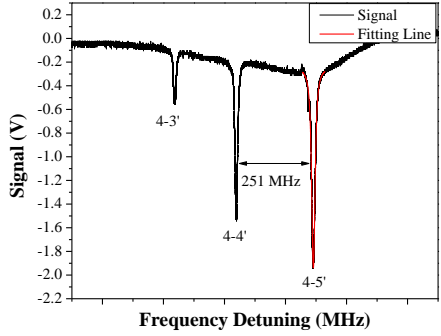

**Fig. 2.** The three hyperfine spectral lines from level $6^2S_{1/2}$ (F=4) to upper levels of the cold cesium atoms. The red line is a Voigt-fitting curve on the 4-5' spectral line.

The total number of cold atoms can be estimated according to the spatial distribution. Inside the integrating sphere, the main effects acting on the cold atoms include the damping force from the cooling laser and gravity. The cooling laser light field is considered as homogeneous and isotropic and so the damping force can be considered uniform. The kinetic energy of a 16-mK cesium atom is about four times the gravitational potential difference in the 40-mm range. Consequently, the cold cesium atoms inside the glass cell have a uniform distribution, and a Boltzmann distribution along the vertical direction. Thus, the density of the cold atoms can be written as

$$n(z) = n_0 \cdot e^{-mgz/k_B T}, \qquad (4)$$

where $n_0$ is the density of the cold atoms at the cell center and m=2.207×$10^{-25}$ kg is the mass of a cesium atom, with $k_B$ the Boltzmann constant and $T$ the temperature of the cold atoms. According to $\Delta N_L$ and Eq. (4), the total number $N$ of the cold atoms at the initial moment is determined to be (2.13±0.01)×$10^9$.

The velocity distribution of the atomic ensemble in the glass cell is too complex to be solved analytically. However, this problem could be simplified if only the cold atoms or thermal atoms are considered. The cold cesium atoms and the thermal background cesium vapor follow two Maxwell distributions, which are $f(v, T)$ and $f(v_b, T_b)$, respectively, determined by the cold atomic temperature $T$ and the ambient temperature $T_b$[17-18]. After blocking the cooling and repumping laser light, the cold cesium atoms inside the glass cell will undergo a specific evolution and eventually be transformed from $f(v, T)$ to $f(v_b, T_b)$. To investigate this evolution process, we measured the absorption of the weak probe laser, whose frequency is scanned at 30 MHz around the 4-5' hyperfine line, at different delay times $\tau$ after the cooling laser and repumping laser are blocked. Within the 30-MHz scanned frequency range, the fluctuation of the 400-MHz background distribution $f(v_b, T_b)$ can be considered negligible compared to the highly resolved absorption spectrum of $f(v, T)$. Thus, the measured signal directly describes the number of the cold cesium atoms populated in the state $6^2S_{1/2}$(F=4). The detailed time sequence of measurements is shown in Fig. 3. In every probing cycle, the cooling and repumping laser light is feed to the cell for 1.5 s, to initialize the atoms to the same starting condition in each measurement. After a certain time $\tau$, the probe laser is turned on and its absorption is detected. For the same $\tau$, the absorption spectrum is plotted by scanning the probe laser frequency in 100 kHz intervals. The measured spectral lines of $\tau$=2 ms, 10 ms, 20 ms, 30 ms, 40 ms, 50 ms, 60 ms, 70 ms, 80 ms, 90 ms, 95 ms, and 100ms are shown in Fig. 3(b). After 100 ms, the spectral line is too weak to be resolved. Thus the lifetime of the cold cesium atoms inside the glass cell is estimated to be approximately 100 ms.

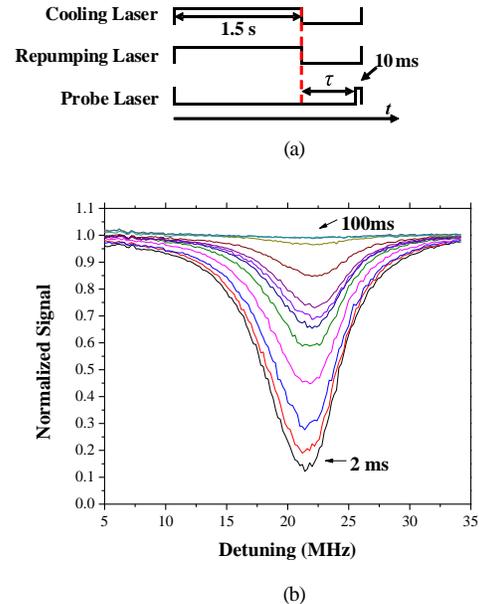

**Fig. 3.** Measurements of the 4-5' hyperfine spectral lines at different time $\tau$: (a) the time sequence for the measurement; (b) from the bottom up, it is the spectral line when $\tau$=2 ms, 10 ms, 20 ms, 30 ms, 40 ms, 50 ms, 60 ms, 70 ms, 80 ms, 90 ms, 95 ms, and 100 ms, respectively.

According to the results shown in Fig. 3(b), the temperature and number of the cold atoms at different $\tau$ are obtained, and plotted in Fig. 4. In Fig. 4(a), we find that within the first 70 ms, the temperature firstly unexpectedly decreases from 16 mK to 9 mK, and then increases rapidly, reaching about 200 mK at 100 ms. The reason of the temperature drop is not entirely understood. We reckon that it should be interpreted using the velocity distribution. After blocking the cooling laser light, the cold atoms with higher velocity within the velocity distribution $f(v, T)$ would first collide with the cell wall. After the collision, these cold atoms are heated to high-speed atoms, or reflected back with a similar velocity, or adsorbed on the wall. As the cold atoms with higher velocity are lost, the global velocity distribution of the optical molasses is dominated by those lower-velocity atoms, which could lead to a drop in temperature. This is consistent with the decrease of cold atom number shown in Fig. 4(b). The lowest temperature of 9 mK appears at 70 ms which equals the time of free falling from the cell center to the bottom. Hence, after 70 ms even the cesium atoms with the lowest velocity around the center experience at least one collision with the cell wall. Moreover, the vertical velocity of about 0.7 m/s caused by the gravity is considerable. This additional velocity leads to more frequent collisions with the wall. Then the global temperature of the optical molasses begins to increase rapidly. Therefore, the temperature evolution can be speculated estimated as the approximate combination of a decrease $e^{-t/\tau_1}$ and an increase $e^{t/\tau_2}$, written as

$$T = T_0 + T_1 e^{-t/\tau_1} + T_2 e^{(t-t_0)/\tau_2}, \quad (5)$$

where $\tau_1$ and $\tau_2$ are respectively the decrease and increase time coefficient, $T_0$ is the base temperature, $t_0$ is the delay time for the increasing process, and $T_1$, $T_2$ are amplitude factors. By fitting on the temperature evolution in Fig. 4(a), we find that $\tau_1=(98\pm7)$ ms and $\tau_2=(5.00\pm0.02)$ ms.

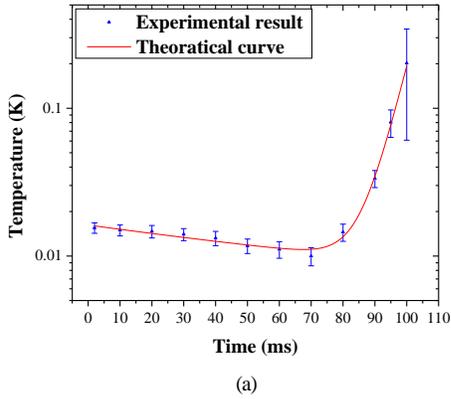

(a)

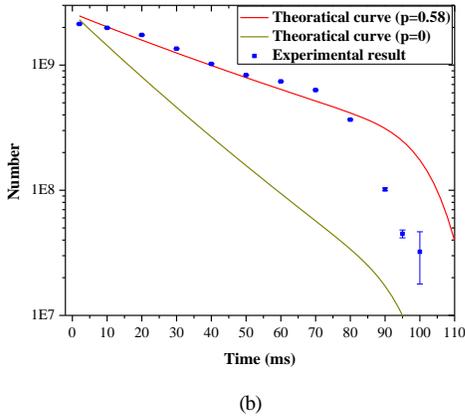

(b)

**Fig. 4.** Two evolutions of the cold cesium atoms: (a) temperature evolution; (b) number evolution.

In Fig. 4(b), the number of cold atoms undergoes a monotonic decrease. After blocking cooling and repumping laser light, the collisional effects account for the majority of the loss of cold atoms. The effects include the collisions of the cold cesium atoms with the other cold cesium atoms (cold-cold), the hot background vapor (cold-hot), and the cell wall (cold-wall). The loss rate $\beta$ of the cold-cold collisions has a relationship with the density of the cold atoms, which is a main concern in a magneto-optical trap with a density larger than $10^9/\text{cm}^3$, and can be interpreted using the Gallagher-Pritchard theoretical model[19]. However, for optical molasses with a density of about $10^7/\text{cm}^3$ inside the integrating sphere, this is too small to be considered. For the cold-hot collisions, the effect produced by the thermal cesium atoms is the main concern because of a much larger cross section compared to the other background thermal atoms[20-21]. The loss of the cold cesium atoms due to the cold-hot collisions can be written as

$$(dN/dt)_{\text{cold-hot}} = -n_b \sigma_{\text{Cs}} v_{\text{brms}} N, \quad (6)$$

where $n_b = P/k_B T_b$ is the density of the thermal cesium atom, $\sigma_{\text{Cs}} = 2\times10^{-13} \text{cm}^2$ is the cesium atomic cross section[21] and $v_{\text{brms}}$ is the thermal root-mean-square velocity. Compared to the cold-wall collisions, the loss rate of the cold-hot collisions is also relatively small. Considering the spatial distribution in Eq. (4), the collisional rate of the cold cesium atoms to the whole cell wall can be expressed as

$$(dN/dt)_{\text{cold-wall}} = \int_S n(z) \cdot \int_0^\infty f(v_r,T) \, dv_r \cdot v_r \, dS, \quad (7)$$

where $v_r$ is the radial velocity of the cold cesium atoms.

The loss of the cold cesium atoms due to the cold-wall collisions is based on the physical processes, where the cold cesium atoms could be adsorbed, inelastically scattered, elastically reflected or changed in state by the surface potential when coming close to the cell wall. It is widely believed that surfaces would heat cold atoms to the ambient temperature in a single collision[6, 10]. However, works on the quantum reflection in past decades are trying to break the traditional thinking and find a surface at room temperature to manipulate cold atoms. Quantum reflection is a reflection of matter waves from the attractive potential, happening when the atom-surface distance $r$ is still beyond several atomic units. Near this region, the interaction can be described by the Casimir-van der Waals potential, which behaves as $-C_4/r^4$ at large distances and as $-C_3/r^3$ at small distances[22-26]. The coefficients $C_3$ and $C_4$ depend on the polarizability property of the cold cesium atoms and on the dielectric property of the cell surface. Quantum reflection is first experimentally observed in the scattering of $^4$He atoms off superfluid helium[27], and subsequently reported in the reflection measurements of atoms from surfaces mediated by the long-range Casimir potential[28-29]. Furthermore, the observations of collisions between hydrogen atoms and bulk helium reveal that the sticking probability should approach zero as $\sqrt{T}$, where $T$ is the temperature of the incoming hydrogen atoms[30]. In 2004, a breakthrough was made by Pasquini *et al.* to demonstrate that "*large quantum reflection probability is not confined to exotic surfaces or extreme angles of*

*incidence*"[22]. Both cold and thermal atom loss rates to the surface could be expressed with an identical equation $dN/dt \propto -(1-R)nSc$, where $R$ is the reflection probability. In their extended work, a remarkable reflection probability of 60% was measured on the pillared silicon surface, indicating a surface with low density help to improve reflectivity[23].

The cesium atom is rarely applied in the quantum reflection because of its large mass and short de Broglie wavelength. In our experiment, although the Casimir-van der Waals potential beyond the close region is not entirely clear on the cell surface, the alkene material coating with a low adsorption energy on the wall may help the cold cesium atoms to meet the reflection requirement. The quantum reflection can keep the population of the cold cesium atoms constant. As the thermal cesium atoms inside an alkene-coated cell can have a rather long decoherence time, the cold atoms with a lower wall-collision rate may survive longer. In order to interpret the measured number evolution due to the cold-wall collision, it may be anticipated that a fraction of the incident cold cesium atoms scattered or reflected by the cell wall remain largely unchanged. We assume the fraction of elastically reflected atoms to be $p$. This means that the cold cesium atoms are heated to the ambient temperature by a single collision with the wall if $p$ is exactly zero. Considering the fraction $p$ and the two-collision effect in Eq. (6) and Eq. (7), the evolution of the cold cesium atoms number inside the glass cell can be written as

$$dN/dt = -(1-p)(dN/dt)_{\text{cold-wall}} + (dN/dt)_{\text{cold-hot}}, \quad (8)$$

where $N$ is the total number of cold atoms. The temperature plays an essential role in the number evolution, which decreases after the temperature evolution changes from decreasing to increasing, as shown in Fig. 4. Because the expression of the temperature evolution, Eq. (4), is rather complex, the differential Eq. (8) cannot be solved analytically. The theoretical curve solved numerically with the fraction $p=0.58$ in Fig. 4(b) is in good agreement with the experimental result before 90 ms, but the discrepancy between the theoretical and experimental results after 90 ms becomes noticeable. We speculate it may be caused by the changing horizontal distribution of the cold atoms. After 90 ms, the spatial distribution of the cold cesium atoms inside the glass cell becomes quite inhomogeneous because of the free fall[10], and follows a Gaussian distribution[6, 9]. Thus the spatial distribution in Eq. (4) should be revised based on more careful measurement. However, restricted by the integrating sphere, it is hard to use a charge-coupled device (CCD) to directly observe the optical molasses. The boundary effect of the numerical solution may also account for the discrepancy. For comparison, the theoretical curve of $p=0$ is also presented in Fig. 4(b). This curve shows a more considerable discrepancy with the experimental data.

The fitting result yields the fraction $p$ to be about 0.58 based on a relatively simple model. Although some factors, such as the exact size of the cold atomic cloud, the real horizontal spatial distribution, the surface potential field, are not included in this model and more precise explorations are needed to confirm an exact $p$ value, the results of this work are still interesting. Whether the cold cesium atoms are all heated to the ambient temperature in a single collision with the cell wall, especially a wall coating, for example, an alkene material remains to be determined. A positive outcome could make a substantial difference to precision measurements.

## 4. Conclusion

We have conducted a detailed study on the temperature and number evolutions of the cold cesium atoms diffusively cooled inside a glass cell. During the 100-ms evolution, the temperature of the cold atoms first gradually decreases because of the temperature gradient, and then rapidly increases. The number evolution is a decrease process. Both the temperature and number evolution can provide helpful guidance to improve the precise instruments based on diffuse laser cooling; for example, atomic clocks, quantum information processing, and nonlinear spectroscopy. The number evolution is interpreted with the help of the temperature evolution and a fraction $p$ of the elastic atomic reflection. The theory is overall in good agreement with the experimental variation. According to the analysis, the fraction $p$ has a nonzero value, which appears to indicate that the cold cesium atoms are not all heated to the ambient temperature in a single collision with the cell wall.

## Acknowledgements

The authors specially acknowledge Dr. Xiong Wei, Dr. Li Yu-Hang, Dr. Xu Zhong-Yang and Chi Xu for their helpful discussions.